\documentclass[twocolumn,showpacs,preprintnumbers,amsmath,amssymb]{revtex4-1}
\usepackage{graphicx}
\usepackage{dcolumn}
\usepackage{bm}
\usepackage{color}
\begin{document}
%
%
%
\title{On the theory of the universal dielectric relaxation}
\author{John Y. Fu}
\date{\today}
\affiliation{Department of Mechanical and Aerospace Engineering, The State University of New York, Buffalo, NY, 14260, USA}
%
%
\begin{abstract}
Dielectric relaxation has been investigated within the framework of a modified mean field theory, in which the dielectric response of an arbitrary condensed matter system to the applied electric field is assumed to consist of two parts, a collective response and a slowly fluctuating response; the former corresponds to the cooperative response of the crystalline or noncrystalline structures composed of the atoms or molecules held together by normal chemical bonds and the latter represents the slow response of the strongly correlated high-temperature structure precursors or a partially ordered nematic phase. These two dielectric responses are not independent of each other but rather constitute a dynamic hierarchy, in which the slowly fluctuating response is constrained by the collective response. It then becomes clear that the dielectric relaxation of the system is actually a specific characteristic relaxation process modulated by the slow relaxation of the nematic phase and its corresponding relaxation relationship should be regarded as the universal dielectric relaxation law. Furthermore, we have shown that seemingly different relaxation relationships, such as the Debye relaxation law, the Cole-Cole equation, the Cole-Davidson equation, the Havriliak-Negami relaxation, the Kohlrausch-Williams-Watts function, Jonscher's universal dielectric relaxation law, etc., are only variants of this universal law under certain circumstances.
\end{abstract}
\maketitle
%
%
In 1913, Debye investigated the anomalous dispersion phenomenon, in which the index of refraction falls with the angular frequency of electromagnetic waves, of a group of dipolar molecules \cite{debye1913}. He treated a dipolar molecule of the group as a sphere immersed in a viscous fluid; under the assumption that the only electric field acting on the molecule is the external field, he used Einstein's theory of the Brownian motion \cite{einstein1905,einstein1906} to tackle the collisions between the molecule and its neighboring molecules in the liquid and then studied the dielectric dispersion of the group. Eventually he formulated the following equation \cite{debye1913},
\begin{equation}
\frac{\varepsilon(\omega)-\varepsilon_{\infty}}{\varepsilon_{s}-\varepsilon_{\infty}}=\frac{1}{1+i\omega\tau_{c}},
\label{debye}
\end{equation}
where $\varepsilon(\omega)$ is the complex permittivity of the group of dipolar molecules and $\omega$ is the angular frequency of the external electric field; $\varepsilon_{\infty}$ and $\varepsilon_{s}$ represent the permittivity at the high frequency limit and the static permittivity of the group, respectively; $\tau_{c}$ is the characteristic relaxation time of the group. The above equation is often called the Debye relaxation law, which represents the dielectric response of rotational dipolar molecules to an alternating external electric field.

Dielectric relaxation phenomena had been extensively investigated long before the Debye relaxation law was proposed. For instance, Kohlrausch introduced the stretched exponential function (it is now also called the Kohlrausch function) to describe the charge relaxation phenomenon in Leiden jars in 1854 \cite{kohlrausch1854,cardona2007}. However, the Debye relaxation law might be the first relaxation relationship derived based on statistical mechanics; thus it has been often used as the starting point for investigating relaxation responses of dielectrics. Unfortunately, numerous experimental studies have demonstrated that the relaxation behavior of a wide range of dielectric materials deviates strongly from the Debye relaxation law. Over the past 100 years, many empirical relaxation laws or relationships, which can be regarded as variants of the Debye relaxation law, have been developed. Among the most important are the Cole-Cole equation (1941-1942) \cite{cole1941,cole1942}, the Cole-Davidson equation (1950-1951) \cite{coledavidson1950,coledavidson1951}, the Havriliak-Negami equation (1966-1967) \cite{hn1966,hn1967}, the Kohlrausch-Williams-Watts function (the Fourier transform of the Kohlrausch function) (1970) \cite{ww1970}, etc. In practice, these empirical relationships work well for certain materials under specific conditions, but not for others.

In 1970s, Jonscher and his co-workers analyzed dielectric properties of many insulating and semiconducting materials; he then suggested that there exists a universal law of dielectric responses \cite{jonscher1974,jonscher1975a,jonscher1975b,jonscher1977,jonscher1995}. Jonscher's work further stimulated scientific curiosity to explore the physical mechanism underlying the universal relaxation phenomenon. It is now well known that relaxation phenomena characterized by different physical quantities, such as strain, permittivity, etc., are very similar in different materials and most relaxation data can be interpreted by two types of experimental fitting functions: the Kohlrausch function, which is written below \cite{kohlrausch1854}
\begin{equation}
f(t)\cong\mathrm{exp}\left[-\left(\frac{t}{\tau_{ks}}\right)^{\beta}\right] \ \ \ \ \ \mbox{$0<\beta<1$}
\label{stretchedfunction0}
\end{equation}
or the Jonscher function given below \cite{jonscher1975b}
\begin{equation}
f(t)\cong\left\{\begin{array}{ll}
             \mu t^{m_{j}} & \ \ \ \ \ \mbox{$0<m_{j}<1$} \\
             \nu t^{n_{j}-1} & \ \ \ \ \ \mbox{$0<n_{j}<1$}
             \label{jonscherfunction0}
             \end{array}
             ,
\right.
\end{equation}
where $t$ represents time in seconds; $\tau_{ks}$, $\mu$, and $\nu$ are assumed to be constant values for a given material.

\begin{figure}[h!]
\begin{center}
\includegraphics[width=1.0\columnwidth]{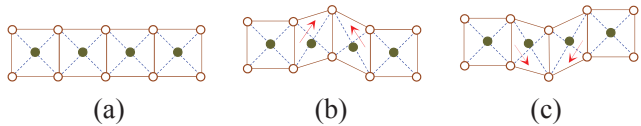}
\vspace{-0.15in}
\caption{Schematic representation of the formation of distorted crystal lattice in paraelectric perovskite systems; (a) undeformed lattice without net dipole moment; (b) and (c) distorted lattice with net dipole moment represented by arrows.}
\vspace{-0.2in}
\end{center}
\end{figure}

Despite having certain ``{\it universal}" relationships, such as the above-mentioned two fitting functions, to describe similar relaxation phenomena in different condensed matter systems, their origins are not entirely clear yet \cite{nondebye1987,jonscher1999,ngai2011}. In this letter, we will attempt to explore the physical mechanism underlying dielectric relaxation phenomena and give a unified interpretation of general relaxation phenomena. We would like to make the conjecture that, if the universal relaxation law does exist, its corresponding mathematical expressions might involve some kinds of ``{\it order parameters}" just like the ones employed in the Landau theory of phase transitions. To make our reasoning and derivation, which will be presented in this letter, more easily understood, we would first discuss the Landau theory and the Landau-Khalatnikov equation in the text that follows.

In 1937, Landau proposed the concept of broken symmetry to explain the ferromagnetic phase transition; he argued that the paramagnetic phase possesses a high degree of symmetry whereas the ferromagnetic phase has a low one; therefore, the phase transition from paramagnetism to ferromagnetism must involve the occurrence of broken symmetry at the critical point. Landau introduced the {\it order parameter}, a unique thermodynamic variable, to demonstrate his broken symmetry concept \cite{landau1937}. For the ferromagnetic phase transition, the order parameter is defined as the spontaneous magnetization $M$. According to Landau's idea, $M$ has the following values: $M=0$, which corresponds to the high degree of symmetry, if the considered material is in the paramagnetic state and $M=\pm \ constant$, which corresponds to the low one, if it is in the ferromagnetic state. In the absence of the external magnetic field and in the vicinity of the critical temperature $T_{c}$, the Landau free energy $F$ per unit volume up to the fourth order of $M$ can be written as \cite{landau1937}
\begin{equation}
F=F_{0}+\frac{1}{2}aM^{2}+\frac{1}{4}bM^{4}+\cdots,
\label{landau1937}
\end{equation}
where $F_{0}$ is the free energy that is independent of $M$; both $a$ and $b$ are coefficients.

In 1954, Landau and Khalatnikov investigated the relaxation behavior of $M$ and formulated the following equation \cite{landau1954}, which is now often called the Landau-Khalatnikov equation.
\begin{equation}
\gamma\frac{dM}{dt}=-\frac{\partial F}{\partial M},
\label{lk1954}
\end{equation}
where $\gamma$ is a kinetic coefficient, which is believed to be independent of temperature \cite{blinc1974,blinov2010}. The Landau-Khalatnikov equation was originally developed to describe the critical slowing down of fluctuation of the order parameter on approaching the critical point \cite{blinc1974}. However, this equation has a more profound physical significance; it can be interpreted as the law of conservation of energy in relaxation processes, i.e., the kinetic energy associated with fluctuation of the order parameter and dissipated during the corresponding relaxation process is equal to the decrease in the Landau free energy. Therefore, this equation can be exploited to analyze the relaxation behavior of the order parameter in general cases.

\begin{figure}[h!]
\begin{center}
\includegraphics[width=1.0\columnwidth]{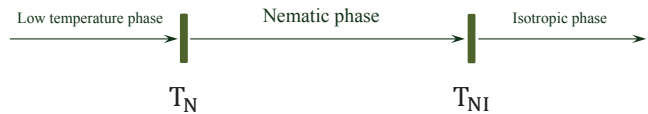}
\caption{Schematic representation of the cooperative behavior spectrum of HTSPs: below temperature $T_{N}$ and above temperature $T_{NI}$, there is no cooperative movement in HTSPs; between $T_{N}$ and $T_{NI}$, HTSPs become strongly correlated and form a nematic phase.}
\vspace{-0.2in}
\end{center}
\end{figure}

We now start to explore whether the universal dielectric relaxation law exists or not. Our conjecture can be further restated as finding a set of fundamental equations, if we do have such a universal law, that can be exploited to derive all of the aforementioned relaxation laws and fitting functions. As usual, we start with the Debye relaxation law. It is generally believed that neglecting mutual interactions between dipolar molecules and their neighboring molecules is the reason that makes the Debye relaxation law deviate from many experimental results. The Debye relaxation law, however, is the model based on statistical mechanics, which is closely related to the Einstein theory of the Brownian motion and the Smoluchowski equation \cite{coffey2006}; like collisions between dipolar molecules and their neighboring molecules, mutual interactions have been implicitly considered in Debye's model and the corresponding induced dipole moments have been assumed to be stochastic in the time domain and randomly distributed in the spatial domain so that the net influence of such mutual interactions is zero. Therefore, in our opinion, the common belief regarding the Debye relaxation law might not be appropriate in most cases. It will become clear later that there exists a specific kind of mutual interactions and its net influence is not zero, which renders the Debye relaxation law inaccurate; the induced dipole moments associated with this mutual interaction are neither purely stochastic in the time domain nor randomly distributed in the spatial domain and their cooperative behavior would eventually alter the dielectric relaxation of a condensed matter system under the perturbation of external electric fields. In the following discussion, for simplicity, we will first consider such mutual interaction in paraelectric perovskites and then extend our conclusion to general dielectric materials.

For a crystalline perovskite, it should have a purely ordered state at the temperature very close to absolute zero and a completely disordered state at its melting point; at temperatures between those two extremes, however, the material must have both ordered and disordered states and the formation of the latter is mainly due to thermal fluctuations (at temperatures near absolute zero, the formation of the disordered state might be determined by the co-operative Jahn-Teller effect \cite{cjte}). Let us consider a perovskite lattice diagrammatically shown in Figs. [1a]-[1c]. At temperatures far below the melting point, there always exists the probability that certain atoms in the normal lattice shown in Fig. [1a] could gain extra kinetic energy from thermal fluctuations to move quasi-permanently away from their original equilibrium positions and distort the original lattice as shown in Figs. [1b] and [1c]. This kind of distortion can introduce the local non-uniform deformation, which breaks the inversion symmetry and induces dipole moments even in paraelectric (non-polar) perovskites \cite{sharma2007}. For simplicity, we define the local disordered structures corresponding to such distorted crystal lattice as high-temperature structure precursors (HTSPs). At new equilibrium positions, HTSPs possess the higher potential due to the induced local strain energy; thus they are often metastable. As shown in Figs. [1b] and [1c], these HTSPs have an equal probability of occurring at different locations (if thermal fluctuations occur randomly) and can be switched from one to the other in either direction under thermal fluctuations; they could also disappear or even ``{\it hop}" to other locations under external perturbations. Therefore, the net influence of HTSPs can be neglected if they behave independently since, under this condition, HTSPs are randomly distributed in the spatial domain and their movements are stochastic in the time domain so that their net effect can be completely averaged out. However, if HTSPs behave cooperatively, the situation could be quite different. Since the formation of HTSPs is closely linked to thermal fluctuations, the quantity of HTSPs is thus dependent upon temperature; as temperature rises, the quantity also increases. At a certain temperature $T_{N}$, the quantity of HTSPs has been raised and reached a threshold or the effective distance between individual HTSPs has been reduced below a critical value so that HTSPs could start to interact with each other, which can be regarded as their cooperative or self-organization behavior, and then form a unique nematic phase. This kind of structural transformation must be regarded as an intrinsic physical phenomenon and the driving force behind it is the competition between energy and entropy. Let us consider a crystalline perovskite containing HTSPs; its Gibbs free energy in the absence of external electric fields can be written as $G=U-TS$. Clearly, the internal energy $U$ will increase due to the increment of the local strain energy and the electric potential generated by HTSPs shown in Fig. 1 (b) or (c). If temperature $T$ remains unchanged or changes slowly, the entropy $S$ must rise to reduce $G$. The simplest way to increase $S$ is that the chemical bonds between the atoms on the distorted lattice sites are partially broken so that the corresponding HTSPs could gain more freedom to rotate and then are oriented along local preferred directions to form a nematic phase, which leads to a decrease in the orientational entropy but an increase in the positional entropy and, eventually, results in a net increase in the total entropy in the considered material. In addition, since the nematic phase has a cylindrical symmetry $\mathrm{D_{\infty h}}$ (no polarity) \cite{blinov2010}, the generated electric energy associated with the distorted lattice shown in Fig. 1 (b) or (c) will decrease when the phase is formed so that $G$ could be further reduced in the considered material. If using $\upsilon$ to represent the local preferred direction and approximately treating individual HTSPs as molecules, we can define the orientational order parameter, $S_{op}$, of the nematic phase as \cite{lubensky2000,degennes1995}
\begin{equation}
S_{op}=\frac{1}{2}\langle3\left(\upsilon^{i}, \vec{n}\right)^{2}-1\rangle=\frac{1}{2}\langle\left(3\mathrm{cos}^{2}\theta^{i}-1\right)\rangle,
\label{orderparameter}
\end{equation}
where $\upsilon^{i}$ is defined as the given preferred direction of the disordered structure located at the position $i$ and $\vec{n}$ is usually called the director that represents the particular direction of the nematic phase; $\theta^{i}$ represents the angle between $\upsilon^{i}$ and $\vec{n}$ at the position $i$; $\langle\ \rangle$ is the mathematical symbol for average. The above-mentioned nematic phase only exists when $0<S_{op}<1$ and, under this condition, all HTSPs behave like a partially ordered liquid. There are two special cases associated with this order parameter: (1) $S_{op}=1$ corresponds to an ideal situation, in which all HTSPs are perfectly aligned along $\vec{n}$; (2) if temperature continues to rise, at a certain point $T=T_{NI}$, although the quantity of HTSPs will continue to increase, any cooperative movement of HTSPs will be destroyed by high-temperature thermal perturbation, which makes HTSPs randomly oriented and then renders $S_{op}=0$. Therefore, when $T>T_{NI}$, all HTSPs behave like a normal liquid, in which individual HTSPs are randomly aligned and there is no preferred direction for their orientation. The spectrum of the cooperative behavior of HTSPs is shown schematically in Fig. [2].

We can now summarize what we have done so far: (1) we first addressed the formation of HTSPs in crystalline dielectrics, which is mainly caused by thermal fluctuations; (2) as temperature rises, the quantity of HTSPs increases and then the effective distance between individual HTSPs decreases. At a certain point, HTSPs start to become strongly correlated and form a nematic phase to reduce the Gibbs free energy of the considered material. It is easy to see that HTSPs become strongly correlated at the cost of the chemical bonds between certain atoms of HTSPs, which are partially broken during the formation of the nematic phase. Therefore, this concept can be easily generalized to polycrystalline and amorphous dielectric materials since the formation of the strongly correlated HTSPs is not dependent on whether material structures are crystalline or not but greatly determined by whether the corresponding chemical bonds are partially broken or not. If considering an amorphous dielectric material, for instance, we can treat it as an effective crystalline material by assuming that it has a crystalline structure with an infinitely long spatial period; then the above-mentioned nematic phase can be formed inside this material for the same reason. In this sense, the only difference between a purely crystalline dielectric and an amorphous dielectric is that the disordered structures are mainly induced by thermal fluctuations in the former, as shown in Fig. 1 (b) or (c), but partially linked to thermal fluctuations in the latter since there are often a large number of inherent disordered structures existing in amorphous materials even at low temperatures. Consequently, our derivation and conclusion drawn from crystalline dielectrics are equally applicable to polycrystalline and amorphous dielectric materials. Here it might be worth briefly discussing the critical temperature $T_{N}$ in both crystalline and amorphous dielectrics. From the above discussion, $T_{N}$ is determined by the quantity of HTSPs that rises above a certain level in the considered material. For a crystalline dielectric, we can find its $T_{N}$ via, for instance, measuring the abrupt change of its refraction index; if somehow we could change the crystalline phase of the dielectric to an amorphous one and then measure its $T_{N}$, we should be able to observe a reduced value of $T_{N}$ due to the existence of inherent disordered structures of the amorphous phase. In other words, compared with crystalline dielectrics, the strongly correlated HTSPs could emerge in amorphous dielectrics at lower temperatures. This conclusion may have a profound significance that could be used to clarify the nature of the so-called Kauzmann's paradox \cite{kauzmann1948}. Further discussion on this topic is omitted in this letter and will be given elsewhere. 

It is also necessary to add some words on the dynamic behavior of the strongly correlated HTSPs or the nematic phase. Notice that the symmetry group of the nematic phase is the continuous rotation group, thus it has uncountable continuum of symmetry elements, which means that its relaxation could be extremely slow \cite{lubensky2000}. It might be interesting to simply compare the dynamic behavior of such a phase with that of glassy materials. Generally speaking, the relaxation of glassy materials is also extremely slow (to be more precise, the relaxation of glassy materials should be slower than that of the nematic phase since there are more disordered structures involved in the former) and their responses, as ensembles of the movement of atoms, to external perturbations are direct. However, the situation of the nematic phase is different; its response to external fields is not directly related to the movement of atoms but largely dependent upon the cooperative movement of HTSPs. When the considered dielectric material is placed under an external field, its equilibrium state will shift from the original one to a new one. At the new equilibrium state, the strongly correlated HTSPs will cooperatively re-adjust themselves and form a new nematic phase to keep the Gibbs free energy to a minimum. Therefore, the dynamics of the nematic phase should be very slow but is different from that of glassy materials.

We now consider the dynamic behavior of the nematic phase of the considered dielectric material under electric fields. In this letter, we assume that the considered material is a dielectric with the simple cubic structure or an amorphous dielectric so that its electric polarization $P$ and electric displacement $D$ can be approximately treated as scalar quantities to simplify our further derivation and discussion. If an ac signal is applied to the considered material, the electric polarization will be induced in its crystalline phases (for crystalline and polycrystalline materials) or other microscopic structures (for amorphous materials), in which the corresponding atoms or molecules are held together by normal chemical bonds, and its nematic phase, respectively; the induced polarization in the former case is denoted by $P$ and the one in the latter case by $P_{n}$. Thus, the formed polarization consists of two parts: $P$ at the normal energy level corresponding to the solid state and $P_{n}$ at the higher energy level corresponding to the partially ordered liquid state. Consequently, the effective polarization inside the considered material could be represented by either of the following two combinations: (a) $P_{eff}=P+P_{n}=(1+k)P$ or (b) $P_{eff}=P-P_{n}=(1-k)P$, and $k$ is defined as
\begin{equation}
k=\frac{k_{0}S_{op}(T-T_{N})}{T_{N}} \ \ \ \mbox{$(T_{N}<T<T_{NI})$},
\label{niphasetransition}
\end{equation}
where $k_{0}$ is a dimensionless coefficient.

In the above-mentioned first case, the interaction between $P$ and $P_{n}$ is neglected; thus, $P_{eff}$ is simply regarded as the superposition of $P$ and $P_{n}$. In the above-mentioned second case, however, the interaction between $P$ and $P_{n}$ cannot be neglected. Since the correlated HTSPs are usually metastable, they can be easily disturbed by the polarization field associated with $P$. According to Le Chatelier's principle, the nematic phase would then undergo a specific structural change to counteract any imposed change by the field \cite{landaulifshitz1980}. Therefore, the effective polarization is defined as $P_{eff}=(1-k)P$ in this case; here the negative sign is due to Le Chatelier's principle. The Gibbs free energy per unit volume of the considered material can then be written as
\begin{equation}
G=U-TS-(1+k)ED=U-TS-(1+k)\frac{\varepsilon_{r}}{\varepsilon_{0}\chi^{2}}P^{2},
\label{dielectricgibbs1}
\end{equation}
which corresponds to the first case, $P_{eff}=(1+k)P$;
\begin{equation}
G=U-TS-(1-k)ED=U-TS-(1-k)\frac{\varepsilon_{r}}{\varepsilon_{0}\chi^{2}}P^{2},
\label{dielectricgibbs2}
\end{equation}
which corresponds to the second case, $P_{eff}=(1-k)P$; here $\varepsilon_{0}$ and $\varepsilon_{r}$ represent the electric permittivity of free space and the relative permittivity, respectively; $\chi$ is defined as the electric susceptibility; $E$ is the applied electric field. The physical meaning of $k$ can be interpreted as follows: it actually represents, in the statistical sense, the fraction of the total potential, which is generated by the cooperative movement of HTSPs. For continuous phase transitions, the theoretical limit for the value of $k$ is $k=0.5$ near the critical point \cite{fulandau}; for dielectric relaxation studies, $k$ is determined by the number of perturbations of external electric fields and may vary from 0 to 1 in different materials, which will be explained in detail later in this letter.

\begin{figure}[h!]
\begin{center}
\includegraphics[width=1.0\columnwidth]{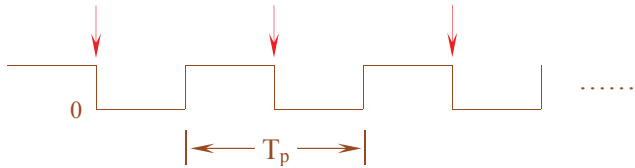}
\caption{Diagrammatic sketch of a testing signal; the characteristic relaxation fully takes place only once within one $T_{p}$, which is represented by vertical arrows; here $T_{p}$ is the period.}
\vspace{-0.2in}
\end{center}
\end{figure}

In the above equations, we have introduced a generalized order parameter, $(1+k)P$ or $(1-k)P$, to define the framework of a modified mean field theory, in which two distinct polarization quantities, $P$ and $P_{n}$, are used to represent the effective response of the considered material to the applied electric field. In conventional mean field theories, the response of the considered material to external perturbations (e.g., temperature, magnetic field, electric field, etc.) is assumed to be an average value; the mutual interactions between individual microscopic particles or structures due to thermal fluctuations in the material are assumed to be randomly distributed in the spatial domain and take place stochastically in the time domain so that the net result of such mutual interactions is zero and can be neglected. In our modified mean field theory, however, such mutual interactions are not treated as random variables but the response of a partially ordered liquid-like phase (nematic phase); in other words, HTSPs become strongly correlated due to such mutual interactions. Since the nematic phase possesses a global symmetry, its response to the applied electric field can be integrated into the Gibbs free energy given by Eqs. (\ref{dielectricgibbs1}) and (\ref{dielectricgibbs2}). Clearly, there is no Landau's broken symmetry concept directly involved in this modified mean field theory; nevertheless, the relaxation of the generalized order parameter can still be investigated via the following equation
\begin{equation}
\gamma\frac{dP}{dt}=\frac{\partial G}{\partial P}.
\label{modifiedlk2012a}
\end{equation}
This equation is the modified Landau-Khalatnikov equation and can be interpreted as follows: the kinetic energy associated with the dynamics of the generalized order parameter and dissipated during the corresponding relaxation process is equal to the decrease in the Gibbs free energy. The difference between this equation (Eq. (\ref{modifiedlk2012a})) and the Landau-Khalatnikov equation (Eq. (\ref{lk1954})) is that there is a negative sign in the right hand side of Eq. (\ref{lk1954}) but none in that of Eq. (\ref{modifiedlk2012a}). This is because the order parameter $M$ does positive contribution to the Landau free energy whereas the generalized order parameter $(1+k)P$ or $(1-k)P$ does negative contribution to the Gibbs free energy.

\begin{figure}[h!]
\begin{center}
\includegraphics[width=1.0\columnwidth]{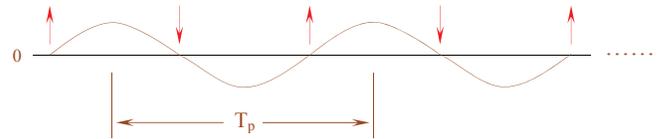}
\caption{Diagrammatic sketch of a cosine signal (testing signal); the characteristic relaxation fully takes place twice within one $T_{p}$, which are represented by vertical arrows; here $T_{p}$ is the period.}
\vspace{-0.2in}
\end{center}
\end{figure}

The modified Landau-Khalatnikov equation only describes a single dielectric relaxation process. In practice, however, the applied electric field exploited to study dielectric relaxation is often a periodic continuous-time signal; many relaxation processes could occur in the considered material during the probing. Thus, in order to understand the relaxation behavior of the considered material, we must study the total of those relaxation processes. For this purpose, we consider a simple testing signal shown schematically in Fig. [3]. This signal is a one-direction testing signal and there is no reversal in it. Let us define its period as $T_{p}$. Notice that the relaxation of the generalized order parameter could be regarded as the ensemble of relaxation of atoms and the theoretical limit of the duration for atomic relaxation is approximately $\tau_{D}=\frac{1}{f_{D}}\approx10^{-13}\mathrm{s}$, where $f_{D}$ is the Debye frequency and $f_{D}\equiv10^{12}\mathrm{Hz}\sim10^{13}\mathrm{Hz}$ \cite{am1976}. If $T_{p}\gg\tau_{D}$, a fully relaxed process would occur only once within one $T_{p}$ as shown in Fig. [3]. Thus, the effective dielectric relaxation, $r_{eff}$, of the considered material under the perturbation of the testing signal can be written as
\begin{equation}
r_{eff}=\frac{1}{N}\displaystyle\sum_{j=1}^{N}r[t+(j-1)T_{p}] \ \ \ (\mbox{$0<t<\tau_{m}$}),
\label{effectiverelaxationt}
\end{equation}
where $r(t)$ is defined as the characteristic relaxation function that represents a single relaxation process fully relaxed between 0 and $\tau_{m}$; $\tau_{m}$ is the modified characteristic relaxation time, which will be defined later, and $\tau_{D}<\tau_{m}\ll T_{p}$; $N$ is the total number of $r(t)$, which is defined as $N=\mathrm{int}(\frac{T_{D}}{T_{p}})$ and here $T_{D}$ is the duration, throughout which the testing signal is applied. The reason we are able to use Eq. (\ref{effectiverelaxationt}) to represent $r_{eff}$ is that we neglect the fatigue behavior of $r(t)$, i.e., each single relaxation process is assumed to be fully relaxed and is independent of other relaxation processes with different time delays. Furthermore, if using $\digamma(\Box)$ to denote the Fourier transform of $\Box$, we can write $r_{eff}$ in the frequency domain as
\begin{eqnarray}
\digamma(r_{eff}) & = & \frac{1}{N}\displaystyle\sum_{j=1}^{N}\digamma(r[t+(j-1)T_{p}]) \nonumber \\
& = & \frac{\digamma(r(t))}{N}\displaystyle\sum_{j=1}^{N}\mathrm{exp}[2\pi i(j-1)T_{p}f] \nonumber \\
& = & \digamma(r(t)) \ \ \ (\mbox{$0<t<\tau_{m}$}),
\label{effectiverelaxationf}
\end{eqnarray}
here $f$ is the frequency of the testing signal and, thus, $T_{p}f=1$. The conclusion drawn from this result would be quite surprised since it shows that the  dielectric relaxation of the considered material is seemingly independent of the testing signal. To take a close look at this surprised result, we further consider a standard testing signal, a cosine signal, for dielectric relaxation studies. Let us also define the period of the cosine signal as $T_{p}$. This signal is a reversed testing signal and has two reversals in one $T_{p}$. If $T_{p}\gg\tau_{D}$, there should be two fully relaxed processes with opposite directions occurring within one $T_{p}$ as shown in Fig. [4]. In this case, $r_{eff}$ can be written as
\begin{eqnarray}
r_{eff} & = & \frac{1}{N}\displaystyle\sum_{j=1}^{N}\{r[t+(j-1)T_{p}] \nonumber \\
& & -r[t+(j-\frac{1}{2})T_{p}]\} \ \ \ (\mbox{$0<t<\tau_{m}$}).
\label{effectiverelaxationt1}
\end{eqnarray}
Similarly, the Fourier transform of $r_{eff}$ is given below
\begin{eqnarray}
\digamma(r_{eff}) & = & \frac{\digamma(r(t))}{N}\displaystyle\sum_{j=1}^{N}\{\mathrm{exp}[2\pi i(j-1)T_{p}f] \nonumber \\
& & -\mathrm{exp}[2\pi i(j-\frac{1}{2})T_{p}f]\} \nonumber \\
& = & 2\digamma(r(t)) \ \ \ (\mbox{$0<t<\tau_{m}$}).
\label{effectiverelaxationf1}
\end{eqnarray}
It is interesting to notice that the effective relaxation processes derived under different testing signals are very similar, i.e., both of them are seemingly independent of testing signals; the only difference between them is in their magnitude values. Therefore, for simplicity, we will choose the signal defined in Fig. [3] as the testing signal in our further discussions. To understand this interesting result, we will derive a specific relaxation formula corresponding to $r(t)$ and then analyze its behavior under different situations in the text that follows.

Case I - the interaction between $P$ and $P_{n}$ is neglected and the generalized order parameter is $P_{eff}=(1+k)P$. In this case, the Gibbs free energy per unit volume, $G$, of the considered material is given by Eq. (\ref{dielectricgibbs1}). Substituting $G$ into Eq. (\ref{modifiedlk2012a}), we get the following result
\begin{equation}
\gamma\frac{dP}{dt}=\frac{\partial G}{\partial P}=-2(1+k)\frac{\varepsilon_{r}}{\varepsilon_{0}\chi^{2}}P.
\label{modifiedlk2012b}
\end{equation}
Solve this equation, we have
\begin{eqnarray}
P & = & P_{s}\mathrm{exp}\left[-\frac{2(1+k)\varepsilon_{r}}{\gamma\varepsilon_{0}\chi_{e}^{2}}t\right] \nonumber \\
& = & P_{s}\mathrm{exp}\left[-\frac{t}{\tau_{m}}\right] \ \ \ (\mbox{$0<t<\tau_{m}$}),
\label{relaxation}
\end{eqnarray}
where $P_{s}$ is defined as the induced electric polarization under static electric fields (since $P_{s}=P|_{t=0}$); $\tau_{m}$ is defined as the modified characteristic relaxation time, which has appeared in Eq. (\ref{effectiverelaxationt}), and $\tau_{m}=\frac{\gamma\varepsilon_{0}\chi_{e}^{2}}{2(1+k)\varepsilon_{r}}$. If we let $k=0$, then we will get another relaxation time, $\tau_{c}=\frac{\gamma\varepsilon_{0}\chi_{e}^{2}}{2\varepsilon_{r}}$. We define $\tau_{c}$ as the characteristic relaxation time of the considered material; the relationship between $\tau_{m}$ and $\tau_{c}$ can be written as
\begin{equation}
\tau_{m}=\frac{\tau_{c}}{1+k}=(1-k_{c})\tau_{c},
\label{tauandmtau1}
\end{equation}
where $k_{c}$ is defined as
\begin{equation}
k_{c}=\frac{k}{1+k}.
\label{kc}
\end{equation}

It might be worth adding a few words about $\tau_{m}$ and $\tau_{c}$. These two parameters are not {\it real relaxation times} but the values at which the relaxed physical quantity (e.g., polarization in our studies) is reduced to $\frac{1}{e}\approx0.368$ times its initial value. For most dielectric studies, it is almost impossible to have the relaxed quantity decayed to $\frac{1}{e}$ times its initial value since the testing signal is usually weak. Therefore, the actual relaxation time, $\tau$, of the relaxed quantity should be $\tau_{D}<\tau<\tau_{m}$ or $\tau_{D}<\tau<\tau_{c}$. We here use $\tau_{m}$ or $\tau_{c}$ instead of $\tau$ for convenience only since it is relatively difficult to determine $\tau$ in practice.

We can also re-write Eq. (\ref{relaxation}) as
\begin{equation}
\chi=\chi_{s}\mathrm{exp}\left[-\frac{t}{\tau_{m}}\right] \ \ \ (\mbox{$0<t<\tau_{m}$})
\label{relaxation1}
\end{equation}
or
\begin{equation}
\varepsilon=\varepsilon_{s}\mathrm{exp}\left[-\frac{t}{\tau_{m}}\right] \ \ \ (\mbox{$0<t<\tau_{m}$}),
\label{relaxation2}
\end{equation}
here $\chi_{s}$ represents the static susceptibility; $\varepsilon$ and $\varepsilon_{s}$, for simplicity, represent the relative permittivity and the static relative permittivity, respectively. Since $\tau_{D}<\tau<\tau_{m}$, there must be a high frequency limit imposed on the dielectric relaxation of the considered material beyond which the material cannot fully respond to the testing signal and its relaxation will start to saturate. Without loss of generality, we define the value of $\varepsilon$ at this limit as $\varepsilon_{\infty}$. Then Eq. (\ref{relaxation2}) will be modified as
\begin{equation}
\frac{\varepsilon-\varepsilon_{\infty}}{\varepsilon_{s}-\varepsilon_{\infty}}=\mathrm{exp}\left[-\frac{t}{\tau_{m}}\right] \ \ \ (\mbox{$0<t<\tau_{m}$}).
\label{relaxation3}
\end{equation}

Now we know that $r(t)=\mathrm{exp}(-\frac{t}{\tau_{m}})$. Thus, the effective dielectric relaxation of the considered material in the time domain can be written as
\begin{equation}
r_{eff}=\frac{1}{N}\displaystyle\sum_{j=1}^{N}\mathrm{exp}\left[-\frac{t}{\tau_{m}}+(j-1)T_{p}\right],
\label{c1effectiverelaxationt}
\end{equation}
where $t$ varies from 0 to $\tau_{m}$. In the frequency domain, according to Eq. (\ref{effectiverelaxationf}), this effective relaxation can be written as
\begin{equation}
\digamma(r_{eff})=\digamma\left[\mathrm{exp}\left(-\frac{t}{\tau_{m}}\right)\right] \ \ \ (\mbox{$0<t<\tau_{m}$}).
\label{c1effectiverelaxationf}
\end{equation}

To take the Fourier transform of the right hand side of Eq. (\ref{c1effectiverelaxationf}), we first assume that $t$ varies from 0 to $\infty$ and then get the following Fourier transform result.
\begin{equation}
\digamma\left[\mathrm{exp}\left(-\frac{t}{\tau_{m}}\right)u(t)\right]=\frac{\tau_{m}}{1+i\omega\tau_{m}},
\label{relaxation4}
\end{equation}
where $\omega=2\pi f$. Since $t$ in Eq. (\ref{c1effectiverelaxationf}) actually varies from 0 to $\tau_{m}$, we have to modify Eq. (\ref{relaxation4}) to get $\digamma(r_{eff})$. If assuming that we squeeze the waveform of the testing signal from $(0,\infty)$ to $(0,\tau_{m})$ in the time domain, we can modify Eq. (\ref{relaxation4}) as
\begin{eqnarray}
\digamma(r_{eff}) & = & \frac{\varepsilon(\omega)-\varepsilon_{\infty}}{\varepsilon_{s}-\varepsilon_{\infty}} \nonumber \\
& = & \frac{1}{\tau_{m}}\digamma\left[\mathrm{exp}\left(-\frac{t}{\tau_{m}}\right)u(t)\right] \nonumber \\
& = & \frac{1}{1+i\omega\tau_{m}}.
\label{relaxation5}
\end{eqnarray}
To justify if this modification is appropriate or not, we can imagine that the waveform must behave like the Dirac delta function if it is squeezed to a point. Now we substitute 0 for $\tau_{m}$ in the above equation and then get $\digamma(r_{eff})=1$, which is exactly the Fourier transform result of the Dirac delta function. Therefore, Eq. (\ref{relaxation5}) represents the effective dielectric relaxation of the considered material in the frequency domain. Furthermore, both Eq. (\ref{effectiverelaxationf}) and Eq. (\ref{relaxation5}) also indicate that $r_{eff}$ could be effectively represented by a stretched exponential function. In fact, Eq. (\ref{c1effectiverelaxationt}) represents not only an average of individual atomic relaxation processes but also a process of time stretching; by adding together each relaxation process occurring in the corresponding time interval $(0,\tau_{m})$, the time variable, $t$, and the time parameter, $\tau_{m}$, associated with the considered dielectric relaxation will be repeatedly expanded from one time interval to another. If $T_{D}$ of the testing signal is much greater than its $T_{p}$, we can assume that both $t$ and $\tau_{m}$ are effectively stretched from $(0,\tau_{m})$ to $(0,\infty)$. Therefore, the effective dielectric relaxation in the time domain can be written as
\begin{equation}
r_{eff}\equiv F_{s}=\mathrm{exp}\left(-\frac{t}{\tau_{sm}}\right) \ \ \ (\mbox{$0<t<\infty$}),
\label{stretchedtimedomain}
\end{equation}
where $F_{s}$ is used to denote the exponential function for convenience; $\tau_{sm}$ and $\tau_{sc}$ are the stretched forms of $\tau_{m}$ and $\tau_{c}$, respectively. Because both $t$ and $\tau_{m}$ are stretched simultaneously, the ratio of $\frac{t}{\tau_{m}}$ in $(0,\tau_{m})$ is equivalent to that of $\frac{t}{\tau_{sm}}$ in $(0,\infty)$. Similarly, $\frac{t}{\tau_{c}}$ in $(0,\tau_{c})$ is equivalent to $\frac{t}{\tau_{sc}}$ in $(0,\infty)$. Thus, the relationship between $\tau_{sm}$ and $\tau_{sc}$ can be written as $\tau_{sm}=\frac{\tau_{sc}}{1+k}=(1-k_{c})\tau_{sc}$. Both $\tau_{sm}$ and $\tau_{sc}$ are special time values at which the relaxed dielectric quantities will decay to $\frac{1}{e}$ times their initial values and, therefore, can be determined in practice. We have to emphasize that $\digamma(r_{eff})$ should not be gotten by directly taking the Fourier transform of Eq. (\ref{stretchedtimedomain}); it should be derived from both Eq. (\ref{effectiverelaxationf}) and Eq. (\ref{relaxation5}). Eq. (\ref{stretchedtimedomain}) is just an effective relaxation relationship in the time domain, which is stretched via the average process defined by Eq. (\ref{c1effectiverelaxationt}). Both $\tau_{c}$ and $\tau_{m}$ are constituted by several real physical quantities, whereas $\tau_{sc}$ and $\tau_{sm}$ are only measured values in practice. Despite the above-mentioned problems, it is still worth deriving effective relaxation relationships in the time domain since they could provide certain unique information about relaxation phenomena, which can be seen when we compare the Cole-Davidson equation and the Kohlrausch function later.

So far we have derived dielectric relaxation relationships in both the time and the frequency domains; these relationships will be further analyzed under the following situations.

Case I (a) - if the strongly correlated HTSPs are neglected, i.e., $k=0$, $\tau_{m}$ will reduce to $\tau_{c}$; then Eq. (\ref{relaxation5}) will reduce to the Debye relaxation law given by Eq. (\ref{debye}). The effective Debye relaxation law in the time domain, $F_{d}$, can be derived from Eq. (\ref{stretchedtimedomain}); the result is written below
\begin{equation}
r_{eff}\equiv F_{d}=\mathrm{exp}\left(-\frac{t}{\tau_{sc}}\right) \ \ \ (\mbox{$0<t<\infty$}).
\label{stretcheddebye}
\end{equation}
Here we can see that, just like the Landau theory of phase transitions \cite{fulandau}, the Debye relaxation law is also a mean field model, in which mutual interactions between dipolar molecules and their neighboring molecules are not neglected as commonly believed but are treated as random variables so that their net results are averaged out. Within the framework of our model, mutual interactions are treated as the response of a partially ordered nematic phase over a wide range of temperature. We believe this is the main factor that intrinsically renders the Debye relaxation law inaccurate in most dielectric materials. In the rest of this letter, we will show, if the strongly correlated HTSPs are considered or $k\neq0$, how other relaxation laws as variants of Eq. (\ref{relaxation5}) can be derived.

Case I (b) - $k\neq0$ but $k$ and $k_{c}$ are very small, then Eq. (\ref{relaxation5}) can be re-written as
\begin{equation}
\frac{\varepsilon(\omega)-\varepsilon_{\infty}}{\varepsilon_{s}-\varepsilon_{\infty}}=\frac{1}{1+i\omega\tau_{m}}=\frac{1}{1+i(1-k_{c})\omega\tau_{c}}.
\label{relaxation6}
\end{equation}
Since $k_{c}\sim0$, $i(1-k_{c})\omega\tau_{c}$ can be simplified as $i(1-k_{c})\omega\tau_{c}\approx1-1+k_{c}+i(1-k_{c})\omega\tau_{c}=1+(1-k_{c})[i\omega\tau_{c}-1]$. Since $\omega\tau_{c}\ll1$ (this is because $\tau_{c}\ll T_{P}$), we can further write $|i\omega\tau_{c}-1|\approx1$. Obviously, $1-k_{c}$ is a real number and $1-k_{c}>0$. Thus, by taking advantage of Eq. (\ref{binomial1}) in Appendix, we have $i(1-k_{c})\omega\tau_{c}\approx(i\omega\tau_{c})^{1-k_{c}}$; then Eq. (\ref{relaxation6}) can be simplified as,
\begin{equation}
\frac{\varepsilon(\omega)-\varepsilon_{\infty}}{\varepsilon_{s}-\varepsilon_{\infty}}=\frac{1}{1+(i\omega\tau_{c})^{1-k_{c}}},
\label{colecole}
\end{equation}
which is the mathematical expression of the Cole-Cole equation \cite{cole1941,cole1942}. We can see that the Cole-Cole equation is only valid when the following conditions are met: (a) there are only tiny amounts of the strongly correlated HTSPs in the considered material because $k$ and $k_{c}$ are very small; (b) the frequency of the testing signal should be low; this is partially due to $\tau_{c}\ll T_{P}$ and there is another factor that could also impose this requirement on the Cole-Cole equation, which will be explained in Case III.

Case I (c) - $k\neq0$ and $k$ and $k_{c}$ are not small. Since $|i\omega\tau_{c}|<1$, using Eq. (\ref{binomial1}) in Appendix, we can re-write Eq. (\ref{relaxation6}) as
\begin{equation}
\frac{\varepsilon(\omega)-\varepsilon_{\infty}}{\varepsilon_{s}-\varepsilon_{\infty}}=\frac{1}{(1+i\omega\tau_{c})^{1-k_{c}}},
\label{coledavidson}
\end{equation}
which is the mathematical expression of the Cole-Davidson equation \cite{coledavidson1950,coledavidson1951}. Obviously, the Cole-Davidson equation is superior to the Cole-Cole equation since the former is unaffected by those severe limitations imposed on the applicability of the latter.

The Cole-Davidson equation is a relaxation law in the frequency domain; it might be interesting to derive its corresponding equation in the time domain. Let us re-write Eq. (\ref{stretchedtimedomain}) as
\begin{equation}
r_{eff}\equiv F_{s}=\mathrm{exp}\left[-(1+k)\frac{t}{\tau_{sc}}\right] \ \ \ (\mbox{$0<t<\infty$}).
\label{cdstretched}
\end{equation}
Multiply both sides of the above equation by $\mathrm{exp}(k)$, we get the following result.
\begin{equation}
F_{cd}=\mathrm{exp}\left[-(1+k)\frac{t}{\tau_{sc}}+k\right] \ \ \ (\mbox{$0<t<\infty$}),
\label{cdstretched1}
\end{equation}
where $F_{cd}=F_{s}\mathrm{exp}(k)$, which is used to represent the effective Cole-Davidson equation in the time domain for convenience. We can write $(1+k)\frac{t}{\tau_{sc}}-k=1+(1+k)\left(\frac{t}{\tau_{sc}}-1\right)$. Obviously, we have $|\frac{t}{\tau_{sc}}-1|<1$. Using Eq. (\ref{binomial1}) in Appendix, we get $(1+k)\frac{t}{\tau_{sc}}-k\approx\left(\frac{t}{\tau_{sc}}\right)^{1+k}$. Then the above equation can be re-written as
\begin{equation}
F_{cd}=\mathrm{exp}\left[-\left(\frac{t}{\tau_{sc}}\right)^{1+k}\right] \ \ \ (\mbox{$0<t<\infty$}).
\label{cdstretched2}
\end{equation}
This equation is the effective Cole-Davidson equation in the time domain. Since $1+k>1$, compared with $F_{d}$ defined by Eq. (\ref{stretcheddebye}), the equation can be regarded as the compressed form of the effective Debye relaxation law in the time domain. This conclusion is very useful and can be used to explain the difference between the Cole-Davidson equation and the Kohlrausch-Williams-Watts function in the time domain later.

Case I (d) - in the Cole-Davidson equation (Eq. (\ref{coledavidson})), if we further assume that there exists $k_{hn}$ and $k_{hn}\sim0$, then $1-k_{hn}\approx1$. We thus have $i\omega\tau_{c}\approx i(1-k_{hn})\omega\tau_{c}$. By using the same approach exploited in deriving the Cole-Cole equation, we can get $i\omega\tau_{c}\approx i(1-k_{hn})\omega\tau_{c}\approx (i\omega\tau_{c})^{1-k_{hn}}$. Substituting this expression into Eq. (\ref{coledavidson}), we have the following relationship:
\begin{equation}
\frac{\varepsilon(\omega)-\varepsilon_{\infty}}{\varepsilon_{s}-\varepsilon_{\infty}}=\frac{1}{[1+(i\omega\tau_{c})^{1-k_{hn}}]^{1-k_{c}}},
\label{hnrelaxation}
\end{equation}
which is the mathematic expression of the Havriliak-Negami equation \cite{hn1966,hn1967}. In this equation, unlike $k_{c}$, $k_{hn}$ does not have real physical meaning; in addition, this equation has two exponents to be determined but, compared with the Cole-Davidson equation, cannot provide a better explanation of dielectric relaxation. Therefore, the Havriliak-Negami equation might not be a particularly useful model.

In Case I, the boundary of validity of relaxation relationships (the Cole-Cole equation, the Cole-Davidson equation, the Havriliak-Negami equation) is fundamentally limited by the assumption, no interaction between $P$ and $P_{n}$. In addition, there is another factor that could significantly alter dielectric relaxation behavior but has not been fully considered in Case I; that is that $k$ is frequency-dependent, which will be discussed in detail in Case III. Because of these limitations, the relaxation relationships in Case I, in general, could only be used to study certain materials under low-frequency perturbations.

Case II - the interaction between $P$ and $P_{n}$ cannot be neglected and the generalized order parameter is $P_{eff}=(1-k)P$. In this case, the Gibbs free energy per unit volume of the considered material is defined by Eq. (\ref{dielectricgibbs2}). Using the same method employed in Case I, we get the following result.
\begin{equation}
\frac{\varepsilon-\varepsilon_{\infty}}{\varepsilon_{s}-\varepsilon_{\infty}}=\mathrm{exp}\left(-\frac{t}{\tau_{n}}\right) \ \ \ (\mbox{$0<t<\tau_{n}$}),
\label{kohlrausch}
\end{equation}
where $\tau_{n}$ is defined as
\begin{equation}
\tau_{n}=\frac{\tau_{c}}{1-k}=(1+k_{s})\tau_{c},
\label{kohlrauschnn}
\end{equation}
and $k_{s}$ is given by
\begin{equation}
k_{s}=\frac{k}{1-k}.
\label{ks}
\end{equation}
Thus, the effective dielectric relaxation of the considered material in the time domain can be written as
\begin{equation}
r_{eff}=\frac{1}{N}\displaystyle\sum_{j=1}^{N}\mathrm{exp}\left[-\frac{t}{\tau_{n}}+(j-1)T_{P}\right],
\label{c2effectiverelaxationt}
\end{equation}
where $t$ varies from 0 to $\tau_{n}$. In the frequency domain, this effective relaxation is given below.
\begin{eqnarray}
\digamma(r_{eff}) & = & \frac{\varepsilon(\omega)-\varepsilon_{\infty}}{\varepsilon_{s}-\varepsilon_{\infty}} \nonumber \\
& = & \frac{1}{1+i\omega\tau_{n}}.
\label{c2effectiverelaxationf}
\end{eqnarray}
Clearly, if $k=0$, $\tau_{n}$ will reduce to $\tau_{c}$ and the above equation will reduce to the Debye relaxation law. For $0<k<1$ and $k_{s}>0$, we re-write the above equation as
\begin{equation}
\frac{\varepsilon(\omega)-\varepsilon_{\infty}}{\varepsilon_{s}-\varepsilon_{\infty}}=\frac{1}{1+i\omega\tau_{n}}=\frac{1}{1+i(1+k_{s})\omega\tau_{c}}.
\label{kwwfunction}
\end{equation}
By using the method exploited in deriving the Cole-Davidson equation in Case I (c), we can modify the above equation and then give the result below.
\begin{equation}
\frac{\varepsilon(\omega)-\varepsilon_{\infty}}{\varepsilon_{s}-\varepsilon_{\infty}}=\frac{1}{(1+i\omega\tau_{c})^{1+k_{s}}}.
\label{kwwfunction1}
\end{equation}

We can also write the corresponding effective dielectric relaxation in the time domain as
\begin{equation}
r_{eff}\equiv F_{k}=\mathrm{exp}\left(-\frac{t}{\tau_{sn}}\right)=\mathrm{exp}\left[-(1-k)\frac{t}{\tau_{sc}}\right],
\label{stretchedinc2}
\end{equation}
where $F_{k}$ is used to denote the effective dielectric relaxation in the time domain for convenience; $\tau_{sn}$ is the stretched form of the modified characteristic relaxation time $\tau_{n}$; $t$ varies from 0 to $\infty$. Similarly, the ratio of $\frac{t}{\tau_{sn}}$ in $(0,\infty)$ is equivalent to that of $\frac{t}{\tau_{n}}$ in $(0,\tau_{n})$; thus the relationship between $\tau_{sn}$ and $\tau_{sc}$ can be written as $\tau_{sn}=\frac{\tau_{sc}}{1-k}=(1+k_{s})\tau_{sc}$. Multiplying both sides of the above equation by $\mathrm{exp}(-k)$, we can further modify it and then get the following result.
\begin{equation}
F_{k}\mathrm{exp}(-k)=\mathrm{exp}\left[-(1-k)\frac{t}{\tau_{sc}}-k\right].
\label{stretchedinc3}
\end{equation}
Making the following algebraic simplification: $(1-k)\frac{t}{\tau_{sc}}+k=1+(1-k)\left(\frac{t}{\tau_{sc}}-1\right)$, we get, via Eq. (\ref{binomial1}) in Appendix, $(1-k)\frac{t}{\tau_{sc}}+k\approx\left(\frac{t}{\tau_{sc}}\right)^{1-k}$. Then the above equation can be re-written as
\begin{equation}
F_{sk}=\mathrm{exp}\left[-\left(\frac{t}{\tau_{sc}}\right)^{1-k}\right] \ \ \ (\mbox{$0<t<\infty$}),
\label{stretchedinc4}
\end{equation}
where $F_{sk}=F_{k}\mathrm{exp}(-k)$. It is very interesting to note that the above equation is the mathematic expression of the Kohlrausch function defined by Eq. (\ref{stretchedfunction0}) \cite{kohlrausch1854}. Thus, the Kohlrausch function actually represents the effective dielectric relaxation in the time domain in Case II. Since $1-k<1$, compared with $F_{d}$ defined by Eq. (\ref{stretcheddebye}), the Kohlrausch function is the stretched form of the effective Debye relaxation law in the time domain.

This important function was extended to study relaxation phenomena in the frequency domain by Williams and Watts in 1970 \cite{ww1970}. Now the Fourier transform of the Kohlrausch function is, thus, often called the Kohlrausch-Williams-Watts function. For simplicity, we use $F_{kww}$ to denote the Kohlrausch-Williams-Watts function, which is written below.
\begin{eqnarray}
F_{kww} & = & \digamma\left(\mathrm{exp}\left[-\left(\frac{t}{\tau_{sc}}\right)^{1-k}\right]\right) \nonumber \\
& \approx & \digamma\left(\mathrm{exp}\left[-(1-k)\frac{t}{\tau_{sc}}-k\right]\right) \nonumber \\
& = & \digamma\left(\mathrm{exp}\left[-\frac{t}{\tau_{sn}}-k\right]\right),
\label{fkww}
\end{eqnarray}
where $t$ varies from 0 to $\infty$. It is obvious that deriving $F_{kww}$ directly from taking the Fourier transform of the Kohlrausch function is not easy. We here try to give a simple form of $F_{kww}$ below.

Considering that $\tau_{sn}$ is not constituted by real physical quantities and the ratio of $\frac{t}{\tau_{sn}}$ in $(0,\infty)$ is equivalent to that of $\frac{t}{\tau_{n}}$ in $(0,\tau_{n})$, we can re-write $F_{kww}$ as
\begin{eqnarray}
F_{kww} & = & \digamma\left(\mathrm{exp}\left[-\frac{t}{\tau_{n}}-k\right]\right) \ \ \ (\mbox{$0<t<\tau_{n}$}) \nonumber \\
& = & \mathrm{exp}(-k)\digamma\left(r(t)\right) \ \ \ (\mbox{$0<t<\tau_{n}$}) \nonumber \\
& = & \frac{C}{(1+i\omega\tau_{c})^{1+k_{s}}},
\label{fkww1}
\end{eqnarray}
where the characteristic relaxation function $r(t)=\mathrm{exp}\left(-\frac{t}{\tau_{n}}\right)$ and $C=\mathrm{exp}(-k)$. Comparing the above equation with Eq. (\ref{coledavidson}), we can see that the Kohlrausch-Williams-Watts function and the Cole-Davidson equation have similar mathematic expressions. Lindsey and Patterson systematically studied these two relaxation relationships; they found that, in the time domain, the Cole-Davidson equation has a sharp long time cutoff while the Kohlrausch function decays exponentially at long times \cite{wwandcd}. Now we know that this is because the effective Cole-Davidson equation is the compressed form of the effective Debye relaxation law while the Kohlrausch function is the stretched one in the time domain.

In practice, the Kohlrausch function may have other forms in different fields. We take a look at a particular example as follows. Using the approximation, $\mathrm{exp}(\Box)\approx1+\Box$, we can simplify Eq. (\ref{stretchedinc2}) as
\begin{equation}
F_{k}=1-(1-k)\frac{t}{\tau_{sc}}.
\label{paretox1}
\end{equation}
Since $|\frac{t}{\tau_{sc}}|<1$, via Eq. (\ref{binomial1}) in Appendix, we can further simplify the above equation as
\begin{equation}
F_{k}=\left(1+\frac{t}{\tau_{sc}}\right)^{-(1-k)},
\label{paretox2}
\end{equation}
which is usually called the Pareto law or the Nutting law (Ref. \cite{paretolaw} and the references cited therein). It is obvious that this law is equivalent to the Kohlrausch function.

\begin{figure}[h!]
\begin{center}
\includegraphics[width=1.0\columnwidth]{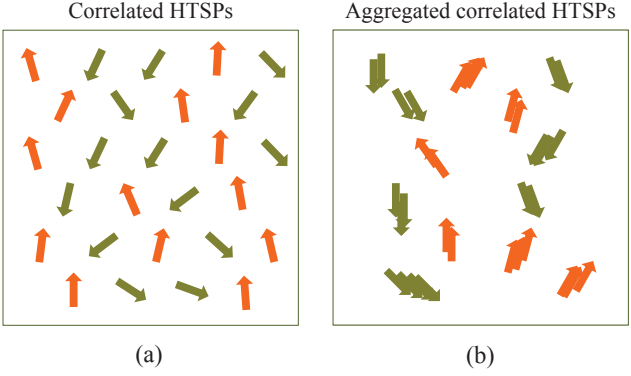}
\caption{Diagrammatic sketches of the formation of nematic phases; (a) the initial nematic phase with the cylindrical symmetry $\mathrm{D_{\infty h}}$ of the correlated HTSPs before exposed to external electric fields; (b) the aggregated nematic phase with $\mathrm{D_{\infty h}}$ of the correlated HTSPs after exposed to external electric fields.}
\vspace{-0.2in}
\end{center}
\end{figure}

In Case II, the interaction between $P$ and $P_{n}$ has been considered. The concept behind this consideration is that the polarization process of crystalline phases or other microscopic structures held together by normal chemical bonds (for simplicity, we will use the {\it normal structures} to represent any microscopic structure held together by normal chemical bonds in the rest of this letter) could be influenced by the nematic phase in the considered material; when the material is placed under an electric field, the induced polarization process (no matter it is the distortional polarization or the orientational polarization) must involve some kind of atomic structural distortion \cite{scaife1998}, which will inevitably disturb the nematic phase; according to Le Chatelier's principle, the nematic phase will undergo a specific structural change to counteract this perturbation and, therefore, the corresponding generalized order parameter must be $P_{eff}=(1-k)P$. In this sense, both the Kohlrausch and the Kohlrausch-Williams-Watts functions are genuine relaxation relationships. Thus, it is not surprised that many relaxation phenomena in different materials can be descried by these two functions.

However, there is one situation that has not been fully considered in both the Kohlrausch and the Kohlrausch-Williams-Watts functions; that is that the parameter $k$ involved in these two function is frequency-dependent, which can significantly alter relaxation behavior in many cases. It was Jonscher who studied relaxation phenomena by implicitly considering $k$ varying under different frequencies and proposed his famous universal dielectric relaxation law \cite{jonscher1974,jonscher1975a,jonscher1975b,jonscher1977,jonscher1995}. Ngai also made critical contributions to the so-called universal relaxation behavior; he developed his relaxation law, the first universal and the second universal models \cite{nondebye1987,ngai2011}, which, in principle, are equivalent to Jonscher's models. In this letter, we only discuss Jonscher's work for simplicity; we will show how his universal relaxation models could be derived within the framework of our model in the text that follows.

Case III - the interaction between $P$ and $P_{n}$ is fully considered and the generalized order parameter could be either $P_{eff}=(1-k)P$ or $P_{eff}=(1+k)P$, which depends on $k$. From our previous discussions, we know that $k$ is closely related to thermal fluctuations. Since the quantity of HTSPs is proportional to temperature, the value of $k$ is also proportional to temperature \cite{fulandau}. Just like thermal fluctuations, external electric fields could also generate HTSPs; the more perturbations of external electric fields, the higher the degree of probability that certain atoms could gain extra kinetic energy to move quasi-permanently away from their original equilibrium positions to form HTSPs in the considered material. Thus, the quantity of HTSPs is proportional to the cycle number $N$ of the testing signal. In other words, the value of $k$ is proportional to the frequency of the testing signal in the considered material if we assume that the corresponding $T_{D}$ is fixed and temperature remains unchanged or changes slowly. Thus, the volume of the nematic phase will increase during the progressive and continuous perturbations of the testing signal, which is shown schematically in Fig. [5].

In our studies, we only consider relaxation phenomena with the steady-state values of $k$ rather than the transient-state ones. Therefore, we can assume that each frequency of the testing signal corresponds to a specific value of $k$. Roughly speaking, if the duration, $T_{D}$, and the amplitude of the testing signal are fixed, the higher the frequency, the larger the value of $k$. This conclusion can be used to explain why the Cole-Cole equation requires that the frequency of the testing signal should be low. Let us consider a dielectric material having a small $k$ value in the absence of external electric fields; after exposed to an applied electric field, if the material still has a small $k$ value, then we can safely say that the frequency of the field should be low, which is the case of the Cole-Cole equation discussed in Case I (b).

In the above discussion, we only consider changes of the value of $k$ and the volume of the nematic phase when the considered material is exposed to an external field. Such changes are illustrated by Fig. [5]; before and after exposed to the field, the nematic phase will undergo a structural transformation shown diagrammatically in Figs. [5a] and [5b]. When the field is removed, a reverse structural transformation from Fig. [5b] to Fig. [5a] in the nematic phase must take place to restore the original equilibrium state, i.e., both $k$ and the nematic phase will relax to their previous value and volume. Here, we have to emphasize that any material relaxation must involve certain fatigue behavior. In our studies, however, the fatigue behavior in dielectric relaxation is neglected; this is because, in practice, the duration of the testing signal is limited and its amplitude is weak.

\begin{figure}[h!]
\begin{center}
\includegraphics[width=1.0\columnwidth]{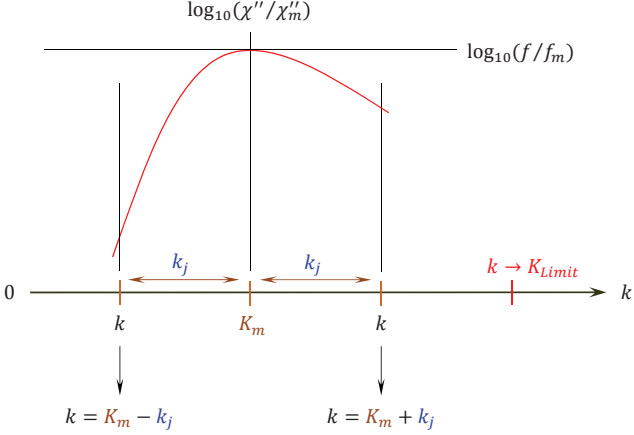}
\caption{Diagrammatic sketches of the values of $k$ before and after the dielectric loss peak: (a) $k=K_{m}-k_{j}$ before the peak and (b) $k=K_{m}+k_{j}$ after the peak; here $f$ is the frequency of the applied electric field; $f_{m}$ and $\chi''_{m}$ represent the frequency value and the imaginary susceptibility at the loss peak; $K_{m}$ is the value of $k$ at the loss peak; $K_{Limit}$ is defined as the theoretical limit of $k$.}
\vspace{-0.2in}
\end{center}
\end{figure}

It is necessary to point out that the strongly correlated HTSPs could also significantly alter the second order phase transition behavior \cite{fulandau}. In that case, $k$ has a different physical meaning. Let us consider a crystalline ferroelectric material; at its critical point $T_{c}$, the old phase collapses but the new phase has not been formed so that the material must be in a completely disordered state. This means $k\rightarrow1$ at the critical point since, in practice, $T_{N}<T_{c}<T_{NI}$ \cite{fulandau}. Thus, at a certain point near $T_{c}$, in the considered material, 50\% of its atoms are in the ordered solid state and the other 50\% are in the partially ordered liquid state, which corresponds to $k\rightarrow0.5$. In the statistical sense, $k\rightarrow0.5$ represents a specific thermodynamic limit; below this limit ($k<0.5$), the material is believed still being a continuum; above this limit ($k>0.5$), the material cannot be regarded as a continuum anymore. So the value of $k$ must be less than 0.5 in second order phase transitions \cite{fulandau}.

In dielectric relaxation studies, however, there is no dramatic structural change involved. Therefore, $k$ is by no means linked to the thermodynamic limit. To some extent, $k$ represents the fraction of the total effective polarization, which is contributed by the nematic phase, negative or positive, of the considered material. From the previous discussion, we know that the atomic structural distortion associated with the polarization process of the normal structures will disturb the nematic phase when the considered material is placed under an external electric field. Thus, the nematic phase, according to Le Chatelier's principle, must undergo a structural change, in which the correlated HTSPs will be aggregated and the volume of the nematic phase will increase as shown schematically in Fig. [5], to counteract this perturbation. The enlarged nematic phase, in turn, will jam the normal structures and try to prevent the further distortion from occurring, or try to hinder the polarization process of the normal structures. In order to complete the polarization process, the considered material needs more energy to overcome the counteractive response of the nematic phase, which, of course, will introduce extra dielectric relaxation loss. In this sense, the generalized order parameter should be $P_{eff}=(1-k)P$. However, the polarization process of the normal structures does not always guarantee being able to perturb the nematic phase. If the frequency of the testing signal is high, on the one hand, the polarization, especially the orientational polarization, of the normal structures could be significantly reduced because the collective atomic movement or dipole rotation in the considered material might not be able to fully keep pace with the change of the testing signal; on the other hand, the volume of the nematic phase will further increase since it is proportional to the cycle number N or the frequency of the testing signal when the corresponding signal duration $T_{D}$ is fixed. Thus, there must exist a critical volume for the nematic phase at a specific frequency, $f_{m}$, beyond which the volume of the nematic phase will be too large to be disturbed. Therefore, when the frequency of the testing signal is greater than $f_{m}$, the nematic phase will not undergo further structural changes to counteract the perturbation exerted by the polarization process of the normal structures but will, independently, respond to the testing signal in the considered material; therefore, the corresponding order parameter will be $P_{eff}=(1+k)P$. It is obvious that, at $f_{m}$, the dielectric relaxation loss will reach the maximum in the considered material; this is partially because, on the high frequency side beyond $f_{m}$, there will be no counteractive response from the nematic phase needed to be overcome in the polarization process of the normal structures. We here denote the value of $k$ at $f_{m}$ as $K_{m}$ for convenience.

Furthermore, under the extreme condition where $f\gg f_{m}$, the polarization of the normal structures will be significantly reduced whereas HTSPs could still be easily polarized even by high-frequency external fields because of their partially ordered liquid phase. Under this situation, the effective polarization of the considered material might be dominated by the contribution from the nematic phase. Thus, $k$ might approach 1 when $f\gg f_{m}$. This is especially the case with polymer materials, in which the orientational polarization dominates. We here use $K_{Limit}\rightarrow1$ to represent the value of $k$ when $f\gg f_{m}$ and obtain $0<k<1$. In the rest of this letter, we will analyze dielectric relaxation phenomena under two situations, $k<K_{m}$ and $k>K_{m}$.

Let us assume that the frequency of the testing signal ranges from the dc region ($f=0$Hz) to the microwave region ($f=3\times10^{8}\sim3\times10^{11}$Hz); under this experimental condition, we have $\omega\tau_{n}\ll1$ in most cases (because the theoretical limit of $\tau_{n}$ is $\tau_{D}\approx10^{-13}$s). Thus, we can re-write Eq. (\ref{kwwfunction}) as
\begin{eqnarray}
\frac{\chi(\omega)-\chi_{\infty}}{\chi_{s}-\chi_{\infty}} & = & \frac{\chi'(\omega)-i\chi''(\omega)-\chi_{\infty}}{\chi_{s}-\chi_{\infty}} \nonumber \\
& = & \frac{\varepsilon(\omega)-\varepsilon_{\infty}}{\varepsilon_{s}-\varepsilon_{\infty}}=\frac{1}{1+i\omega\tau_{n}} \nonumber \\
& = &\frac{1-i\omega\tau_{n}}{1+(\omega\tau_{n})^{2}}\approx1-i\omega\tau_{n},
\label{jonscher}
\end{eqnarray}
where $\chi_{\infty}$ is the susceptibility at the high frequency limit; $\chi'(\omega)$ and $\chi''(\omega)$ are the real and the imaginary parts of $\chi(\omega)$, respectively. Before having further discussions, we need to briefly review Jonscher's universal dielectric relaxation law below.

Within the same frequency range as the above-mentioned, Jonscher studied many dielectric and semiconducting materials and then proposed the following empirical relationships in the frequency domain. \cite{jonscher1974,jonscher1975a,jonscher1975b,jonscher1977,jonscher1995}.
\begin{equation}
\left\{\begin{array}{ll}
             \chi''(\omega)\propto \omega^{m_{j}} & \ \ \ \ \ \mbox{$0<m_{j}<1$ and $f<f_{m}$} \\
             \chi(\omega)\propto (i\omega)^{n_{j}-1} & \ \ \ \ \ \mbox{$0<n_{j}<1$ and $f>f_{m}$}
             \label{jonscheruniversal0}
             \end{array}
             ,
\right.
\end{equation}
which is now often called the universal dielectric relaxation law or Jonscher's universal relaxation law; in the time domain, this law has the mathematical expressions defined by Eq. (\ref{jonscherfunction0}). The above relationships demonstrate that relaxation properties of dielectric materials on both sides of their loss peaks are different, which could explain why the observed loss peaks in most materials are asymmetric in a log $\omega$ plot, being steeper on the left side ($f<f_{m}$) than on the right side ($f>f_{m}$) of the peaks. This feature is shown diagrammatically in Fig. [6].

In following discussions, we will show that Jonscher's universal relaxation law just represents two special cases of a fundamental dielectric relaxation and the asymmetric loss peaks are actually caused by the Kohlrausch-Williams-Watts type relaxation and the Cole-Davidson type relaxation on both sides of the peaks, respectively.

Case III (a) - the interaction between $P$ and $P_{n}$ is fully considered and the generalized order parameter is $P_{eff}=(1-k)P$ under the assumption that $0<k<K_{m}$. In this case, the frequency of the testing signal is less than $f_{m}$. By taking advantage of Eq. (\ref{kohlrauschnn}), we can write $\tau_{n}$ as
\begin{equation}
\tau_{n}=\frac{\tau_{c}}{1-k}=\frac{\tau_{c}}{1-K_{m}+k_{j}},
\label{tonschertnn}
\end{equation}
where $k=K_{m}-k_{j}$ and $k_{j}$ is a parameter defined in Fig. [6]. We can also write $\tau_{sn}$ as
\begin{equation}
\tau_{sn}=\frac{\tau_{sc}}{1-k}=\frac{\tau_{sc}}{1-K_{m}+k_{j}}.
\label{tonschertnne0}
\end{equation}

By using Eqs. (\ref{kwwfunction1}) and (\ref{stretchedinc4}), we can write the effective relaxation relationships of Case III (a) as
\begin{eqnarray}
\frac{\chi(\omega)-\chi_{\infty}}{\chi_{s}-\chi_{\infty}} & = & \frac{\chi'(\omega)-i\chi''(\omega)-\chi_{\infty}}{\chi_{s}-\chi_{\infty}} \nonumber \\
& = & \frac{1}{(1+i\omega\tau_{c})^{1+k_{s}}}
\label{caseiiiaf}
\end{eqnarray}
and
\begin{equation}
F_{skj}=\mathrm{exp}\left[-\left(\frac{t}{\tau_{sc}}\right)^{1-k}\right] \ \ \ (\mbox{$0<t<\infty$}),
\label{caseiiiat}
\end{equation}
where $k_{s}=\frac{k}{1-k}=\frac{K_{m}-k_{j}}{1-K_{m}+k_{j}}$ and $F_{skj}=F_{k}\mathrm{exp}(-k)$. Now we try to derive the simplified expressions of the above relationships. Substituting $\tau_{n}$ into Eq. (\ref{jonscher}), we have
\begin{eqnarray}
\frac{\chi(\omega)-\chi_{\infty}}{\chi_{s}-\chi_{\infty}} & = & \frac{\chi'(\omega)-i\chi''(\omega)-\chi_{\infty}}{\chi_{s}-\chi_{\infty}} \nonumber \\
& \approx & 1-i\omega\tau_{n} \nonumber \\
& = & 1-i\omega\frac{\tau_{c}}{1-K_{m}+k_{j}} \nonumber \\
& = & 1-\frac{im\omega\tau_{c}}{1-K_{m}},
\label{jonscher2}
\end{eqnarray}
where $m$ is defined as
\begin{equation}
m=1-\frac{k_{j}}{1-K_{m}+k_{j}}=\frac{1-K_{m}}{1-k},
\label{mjonscherm}
\end{equation}
which shows that $0<m<1$. The term $m\omega\tau_{c}$ can be written as $m\omega\tau_{c}=1+m(\omega\tau_{c}-1)+m-1$. Since $|\omega\tau_{c}-1|<1$, by using Eq. (\ref{binomial1}) in Appendix, we can simplify $m\omega\tau_{c}$ as $m\omega\tau_{c}\approx(\omega\tau_{c})^{m}+m-1$. Substituting this result into Eq. (\ref{jonscher2}), we then have
\begin{equation}
\frac{\chi'(\omega)-i\chi''(\omega)-\chi_{\infty}}{\chi_{s}-\chi_{\infty}}=1-\frac{i\left[(\omega\tau_{c})^{m}+m-1\right]}{1-K_{m}}.
\label{jonscher3}
\end{equation}
If neglecting the dc terms in the above equation, we can get the following directly proportional relationship
\begin{equation}
\chi''(\omega)\propto\omega^{m} \ \ \ \ \ \mbox{$0<m<1$},
\label{jonscher4}
\end{equation}
which is the first mathematical expression of Jonscher's universal dielectric relaxation law defined by Eq. (\ref{jonscheruniversal0}).

Now we consider the effective relaxation in the time domain under this situation. Using the approximation, $\mathrm{exp}(\Box)\approx1+\Box$, we can re-write Eq. (\ref{stretchedinc2}) as
\begin{eqnarray}
F_{k}=\frac{F_{skj}}{\mathrm{exp}(-k)} & = & \mathrm{exp}\left(-\frac{t}{\tau_{sn}}\right)\approx1-\frac{t}{\tau_{sn}} \nonumber \\
& = & 1-(1-k)\frac{t}{\tau_{sc}} \nonumber \\
& = & 1-(1-K_{m}+k_{j})\frac{t}{\tau_{sc}} \nonumber \\
& = & C_{j}\left(\frac{1}{C_{j}}-\frac{t_{j}}{1-K_{m}+k_{j}}\right) \nonumber \\
& = & C_{j}\left(\frac{1}{C_{j}}-\frac{mt_{j}}{1-K_{m}}\right),
\label{jonschertime1}
\end{eqnarray}
where $C_{j}=(1-K_{m}+k_{j})^{2}=(1-k)^{2}$ and $t_{j}=\frac{t}{t_{sc}}$. Taking advantage of the method used to derive Eq. (\ref{jonscher3}), we can simplify the above equation as
\begin{eqnarray}
\frac{F_{skj}}{\mathrm{exp}(-k)} & \approx & C_{j}\left(\frac{1}{C_{j}}-\frac{mt_{j}}{1-K_{m}}\right) \nonumber \\
& \approx & C_{j}\left(\frac{1}{C_{j}}-\frac{t^{m}_{j}+m-1}{1-K_{m}}\right).
\label{jonschertime2}
\end{eqnarray}
Similarly, if neglecting the dc terms in the above equation, we can get the following relationship in the time domain
\begin{equation}
F_{skj}\cong\mu t^{m}_{j} \ \ \ \ \ \mbox{$0<m<1$},
\label{jonschertime3}
\end{equation}
where $\mu=-\frac{C_{j}\mathrm{exp}(-k)}{1-K_{m}}$. Obviously, this equation is the first mathematical expression of the Jonscher function defined by Eq. (\ref{jonscherfunction0}).

Case III (b) - the interaction between $P$ and $P_{n}$ is fully considered and the generalized order parameter is $P_{eff}=(1+k)P$. In this case, the frequency of the testing signal is greater than $f_{m}$ and $K_{m}<k<K_{Limit}$; the volume of the nematic phase is too large to be disturbed so that the total effective polarization should be the superposition of $P$ and $P_{n}$ in the considered material. By taking advantage of Eq. (\ref{tauandmtau1}), we can write $\tau_{n}$ as
\begin{equation}
\tau_{n}=\frac{\tau_{c}}{1+k}=\frac{\tau_{c}}{1+K_{m}+k_{j}},
\label{tonschertnn1}
\end{equation}
where $k=K_{m}+k_{j}$. We can also write $\tau_{sn}$ as
\begin{equation}
\tau_{sn}=\frac{\tau_{sc}}{1+k}=\frac{\tau_{sc}}{1+K_{m}+k_{j}}.
\label{tonschertnne1}
\end{equation}
Similarly, by taking advantage of Eqs. (\ref{coledavidson}) and (\ref{cdstretched2}), we can write the effective relaxation relationships of Case III (b) as
\begin{eqnarray}
\frac{\chi(\omega)-\chi_{\infty}}{\chi_{s}-\chi_{\infty}} & = & \frac{\chi'(\omega)-i\chi''(\omega)-\chi_{\infty}}{\chi_{s}-\chi_{\infty}} \nonumber \\
& = & \frac{1}{(1+i\omega\tau_{c})^{1-k_{c}}}
\label{caseiiibf}
\end{eqnarray}
and
\begin{equation}
F_{cdj}=\mathrm{exp}\left[-\left(\frac{t}{\tau_{sc}}\right)^{1+k}\right] \ \ \ (\mbox{$0<t<\infty$}),
\label{caseiiibt}
\end{equation}
where $k_{c}=\frac{k}{1+k}=\frac{K_{m}+k_{j}}{1+K_{m}+k_{j}}$ and $F_{cdj}=F_{s}\mathrm{exp}(k)$. It is not surprised that both Eq. (\ref{coledavidson}) and Eq. (\ref{caseiiibf}) are the same in mathematical expressions in the frequency domain and both Eq. (\ref{cdstretched2}) and Eq. (\ref{caseiiibt}) are also the same in the time domain. But the physical mechanisms behind these formulas are different. In Case I (c), the interaction between $P$ and $P_{n}$ is neglected, whereas it is fully considered here; it is because the volume of the nematic phase becomes so large that the relaxation behavior of Case III (b) is forced to deviate from the one given in Case III (a). However, for the sake of simplicity, we still call the relaxation defined by Eqs. (\ref{caseiiibf}) and (\ref{caseiiibt}) the Cole-Davidson type relaxation.

Once again, we try to derive the simplified expressions of the relaxation relationships of Case III (b). Substituting $\tau_{n}$ into Eq. (\ref{jonscher}), we have
\begin{eqnarray}
\frac{\chi(\omega)-\chi_{\infty}}{\chi_{s}-\chi_{\infty}} & \approx & 1-i\omega\frac{\tau_{c}}{1+K_{m}+k_{j}} \nonumber \\
& = & 1+\frac{i(n-1)\omega\tau_{c}}{1-K_{m}-k_{j}} \nonumber \\
& = & \frac{1+i(n-1)\omega\tau_{c}-k}{1-k},
\label{jonscher5}
\end{eqnarray}
where $n$ is defined as
\begin{equation}
n=\frac{2(K_{m}+k_{j})}{1+K_{m}+k_{j}}=\frac{2k}{1+k},
\label{njonschern}
\end{equation}
which shows that $0<n<1$. The term $1+i(n-1)\omega\tau_{c}$ can be written as $1+i(n-1)\omega\tau_{c}=1+(n-1)(i\omega\tau_{c}-1)+(n-1)$. Since $\omega\tau_{c}\ll1$ (this is because, in Case III, we only consider the testing signals with frequencies not beyond the microwave region), $|i\omega\tau_{c}-1|\approx1$. It is obvious that $i\omega\tau_{c}-1\neq-1$ and $n-1>-1$. Thus, using Eq. (\ref{binomial1}) in Appendix, we obtain $1+i(n-1)\omega\tau_{c}\approx(i\omega\tau_{c})^{n-1}+n-1$. Substituting this result into Eq. (\ref{jonscher5}), we then have
\begin{equation}
\frac{\chi(\omega)-\chi_{\infty}}{\chi_{s}-\chi_{\infty}}=\frac{(i\omega\tau_{c})^{n-1}+n-k-1}{1-k}.
\label{jonscher6}
\end{equation}
If neglecting the dc terms in the above equation, we obtain the following directly proportional relationship
\begin{equation}
\chi(\omega)\propto(i\omega)^{n-1} \ \ \ \ \ \mbox{$0<n<1$},
\label{jonscher7}
\end{equation}
which is the second mathematical expression of Jonscher's universal dielectric relaxation law defined by Eq. (\ref{jonscheruniversal0}).

Furthermore, substituting $\tau_{sn}$ into Eq. (\ref{jonschertime1}) and replacing $F_{k}$ and $F_{skj}$ with $F_{s}$ and $F_{cdj}$, respectively, we then obtain
\begin{eqnarray}
F_{s}=\frac{F_{cdj}}{\mathrm{exp}(k)} & \approx & 1-(1+k)\frac{t}{\tau_{sc}} \nonumber \\
& = & 1-(1+K_{m}+k_{j})\frac{t}{\tau_{sc}} \nonumber \\
& = & C_{jj}\left(\frac{1}{C_{jj}}-\frac{t_{j}}{1+K_{m}+k_{j}}\right) \nonumber \\
& = & C_{jj}\left(\frac{1}{C_{jj}}+\frac{(n-1)t_{j}}{1-k}\right),
\label{jonschertime3}
\end{eqnarray}
where $C_{jj}=(1+K_{m}+k_{j})^{2}=(1+k)^{2}$. Using the method exploited to derive Eq. (\ref{jonscher6}), we can simplify the above equation as
\begin{eqnarray}
\frac{F_{cdj}}{\mathrm{exp}(k)} & \approx & C_{jj}\left(\frac{1}{C_{jj}}+\frac{(n-1)t_{j}}{1-k}\right) \nonumber \\
& \approx & C_{jj}\left(\frac{1}{C_{jj}}+\frac{t^{n-1}_{j}+n-2}{1-k}\right).
\label{jonschertime4}
\end{eqnarray}
Similarly, if neglecting the dc terms in the above equation, we obtain the following relationship in the time domain
\begin{equation}
F_{cdj}\cong\nu t^{n-1}_{j} \ \ \ \ \ \mbox{$0<n<1$},
\label{jonschertime3}
\end{equation}
where $\nu=\frac{C_{jj}\mathrm{exp}(k)}{1-k}$. It is obvious that this equation is the second mathematical expression of the Jonscher function defined by Eq. (\ref{jonscherfunction0}).

In view of what have been derived and discussed above, it is clear that what Jonscher's universal dielectric relaxation law describes are just two special cases of the fundamental dielectric relaxation given by Eq. (\ref{c2effectiverelaxationf}), which could be greatly altered by the evolution of the volume of the nematic phase. On the low frequency side ($f<f_{m}$) of the loss peak, the volume of the nematic phase is smaller than the critical volume and, thus, the relaxation follows the Kohlrausch-Williams-Watts relaxation. On the high frequency side ($f>f_{m}$) of the loss peak, however, the volume is larger than the critical volume and, thus, the relaxation follows the Cole-Davidson equation. This is why $\chi''(\omega)$ is asymmetric in a log $\omega$ plot for most dielectric materials. Now it is also clear why the relaxation relationships defined by Eqs. (\ref{effectiverelaxationf}) and (\ref{effectiverelaxationf1}), which are seemingly independent of external fields, could eventually evolve into the ones that are obviously field-dependent. This is because the real relaxation of an arbitrary dielectric material is always modulated by the slow relaxation of its nematic phase.

So far we have only considered the relaxation behavior in the situation where temperature remains unchanged and the evolution of the nematic phase and the value of $k$ are completely determined by external electric fields. From our previous discussions, we know that, in the absence of external fields, the quantity of HTSPs and the formation of the nematic phase are directly related to temperature. Consequently, the value of $k$ is proportional to temperature \cite{fulandau}. Thus, it should be interesting to think about the relaxation behavior in another situation where temperature is continuously changed and the testing signal is a fixed low-frequency perturbation of small amplitude. This means that the change of the nematic phase and $k$ in such a relaxation process is totally dependent on temperature. This kind of temperature-dependent relaxation will be discussed in the text that follows.

Case IV - in this case, the evolution of the nematic phase and the variation of $k$ are governed by temperature changes. Similarly, there must exist a critical volume for the nematic phase at a specific temperature, $T_{m}$ ($T_{N}<T_{m}<T_{NI}$); when the testing temperature is above $T_{m}$, the volume of the nematic phase will be too large to be disturbed by the polarization process of the normal structures in the considered material under the perturbation of the testing signal. Thus, using the reasoning exploited in Case III, the generalized order parameter here can be defined as either $P_{eff}=(1-k)P$ when $T<T_{m}$ or $P_{eff}=(1+k)P$ when $T>T_{m}$. Then, we can simply exploit the previous derivation to obtain the relaxation relationships on both sides of $T_{m}$, respectively.

Case IV (a) - the generalized order parameter is $P_{eff}=(1-k)P$ and $T<T_{m}$. By taking advantage of the method used to derive Eqs. (\ref{caseiiiaf}) and (\ref{caseiiiat}), we can write the effective relaxation relationships of Case IV (a) as
\begin{equation}
\frac{\chi(\omega)-\chi_{\infty}}{\chi_{s}-\chi_{\infty}}=\frac{1}{(1+i\omega\tau_{c})^{1+k_{s}}} \ \ \ (\mbox{$T<T_{m}$})
\label{caseivaf}
\end{equation}
and
\begin{equation}
F_{t}=\mathrm{exp}\left[-\left(\frac{t}{\tau_{sc}}\right)^{1-k}\right] \ \ \ (\mbox{$T<T_{m}$}),
\label{caseivat}
\end{equation}
where $t$ varies from 0 to $\infty$ and $k_{s}$ is defined by Eq. (\ref{ks}); here $F_{t}$ is just used to denote the exponential function of Case IV for convenience. These two equations indicate that, when $T<T_{m}$, the corresponding dielectric relaxation behavior obeys the Kohlrausch function (the Kohlrausch-Williams-Watts in the frequency domain).

Case IV (b) - the generalized order parameter is $P_{eff}=(1+k)P$ and $T>T_{m}$. Similarly, by using the method exploited to derive Eqs. (\ref{caseiiibf}) and (\ref{caseiiibt}), we can write the effective relaxation relationships of Case IV (b) as
\begin{equation}
\frac{\chi(\omega)-\chi_{\infty}}{\chi_{s}-\chi_{\infty}}=\frac{1}{(1+i\omega\tau_{c})^{1-k_{c}}} \ \ \ (\mbox{$T>T_{m}$})
\label{caseivbf}
\end{equation}
and
\begin{equation}
F_{t}=\mathrm{exp}\left[-\left(\frac{t}{\tau_{sc}}\right)^{1+k}\right] \ \ \ (\mbox{$T>T_{m}$}),
\label{caseivbt}
\end{equation}
where $t$ varies from 0 to $\infty$ and $k_{c}$ is defined by Eq. (\ref{kc}). These two equations demonstrate that, when $T>T_{m}$, the corresponding dielectric relaxation behavior obeys the Cole-Davidson equation.

Perhaps, the most distinguishing feature of relaxation processes of Case IV is that, when $T<T_{m}$, the corresponding relaxation is the stretched exponential function whereas, when $T>T_{m}$, it becomes the compressed one in the time domain. This feature might be extremely useful in phase transition and structural transformation studies. In addition, a Jonscher's universal relaxation type law can also be derived in this case; the only difference is that $K_{m}$ here has nothing to do with the frequency of the testing signal but is mainly governed by temperature and represents the value of $k$ at $T_{m}$. However, compared with Eqs. (\ref{caseivaf}-\ref{caseivbt}), this kind of relaxation law cannot provide more useful information about temperature-dependent relaxation phenomena. Thus its derivation is omitted here.

It might be worth pointing out that, though drawn from dielectric relaxation, the derivation and the conclusions given here should apply equally to other relaxation phenomena. For instance, for a solid material undergoing deformation, one can arrive at its structural relaxation formulas having the same mathematical expressions as the ones derived in this letter by assuming that the generalized order parameter is formed by its induced strain quantities.

Concluding remarks - in this letter, we have demonstrated that, over a wide range of temperature, dielectric materials possess two microscopic structures, the normal structures, i.e., the crystalline or noncrystalline structures composed of the atoms or molecules held together by normal chemical bonds, and the nematic phase (the partially ordered liquid-like phase) or the strongly correlated HTSPs. When a dielectric material is placed under an applied electric field, its normal structures will give a collective dielectric response at the normal energy level and its nematic phase will present a slowly fluctuating one at the higher energy level. It is obvious that the normal structures and the nematic phase are not independent of each other. At absolute zero and in the absence of external fields, atomic movement is frozen and the nematic phase does not exist in the considered material. As temperature rises, the nematic phase starts to emerge due to thermal fluctuations (or the co-operative Jahn-Teller effect at temperatures near absolute zero). Now if we apply an electric field to the considered material, the volume of its nematic phase will increase, due to the perturbation of the field, at the cost of the volume of its normal structures; the increase of the former corresponds to the decrease of the latter. In this sense, the nematic phase is constrained by the normal structures. Therefore, the dielectric responses of the normal structures and the nematic phase constitute a dynamic hierarchy, in which the slowly fluctuating response is also constrained by the collective response. This suggests that there does exist a universal dielectric relaxation process, which is actually the characteristic relaxation of the normal structures modulated by the slow relaxation of the nematic phase in the considered material; the corresponding relaxation relationship can then be regarded as the universal dielectric relaxation law and other relaxation relationships are only variants of this universal law under different situations.

\begin{center}
\textbf{\small ACKNOWLEDGMENT}
\end{center}
The research presented here was sponsored by the State University of New York at Buffalo. The author is deeply indebted to Professor Yulian Vysochanskii of Uzhgorod National University of Ukraine for stimulating discussions during the course of this work.

\vspace{0.3in}

\appendix

\section{Binomial series expansion}

For a function $f(x)=(1+x)^{\xi}$, its binomial series expansion can be written as
\begin{eqnarray}
(1+x)^{\xi} & = & 1+\xi x+\frac{\xi(\xi-1)}{2!}x^{2}+\frac{\xi(\xi-1)(\xi-2)}{3!}x^{3}+ \nonumber \\
& & \cdots+\frac{\xi(\xi-1)\cdots(\xi-n+1)}{n!}x^{n}+\cdots. \nonumber
\end{eqnarray}
If $|x|<1$, the above series will converge absolutely for any complex number $\xi$; if $|x|=1$, it will converge absolutely if and only if either $\mathrm{Re}(\xi)>0$ or $\xi=0$; if $|x|=1$ and $x\neq-1$, it will converge if and only if $\mathrm{Re}(\xi)>-1$. Under these conditions, the above series can be simplified as
\[
(1+x)^{\xi}\approx1+\xi x. \qquad \tag{A1}\label{binomial1}
\]
\end{document}